\documentclass[aps,prb,twocolumn,floatfix,amssymb,amsmath,showpacs]{revtex4}

\usepackage{graphicx,color}

\begin{document}

\title{Spin polarized photocurrent from quantum dots}

\author{J.~M.~Villas-B\^{o}as}

\author{Sergio E.~Ulloa}

\author{A.~O.~Govorov}
\affiliation{Department of Physics and Astronomy, and Nanoscale and Quantum
Phenomena Institute, Ohio University, Athens, Ohio 45701-2979}

\date{\today}

\begin{abstract}
In this paper we show that it is possible to switch the spin
polarization of the photocurrent signal obtained from a single
self-assembled quantum dot photodiode under the effect of
elliptically polarized light by just increasing the light intensity.
In the nonlinear mechanism treated here, intense elliptically
polarized light creates an effective exchange interaction between
the exciton spin states through the biexciton state. This effect can
be used as a dynamical switch to invert the spin-polarization of the
extracted photocurrent.  We further show that the effect persists in
realistic ensembles of dots, making this a powerful technique to
dynamically generate spin-polarized electrons.
\end{abstract}

\maketitle

Classical information processing has been based on charge currents
in electronic devices. In the search for quantum alternatives, {\em
spin currents} appear very promising, as typical decoherence times
for the spin are much longer than for electron charge. Spin currents
are the foundation for ``spintronics," a concept that researchers
hope will give additional functionalities and result in improved
electronic devices. The spin degree of freedom also constitutes
natural qubits, the basic unit in quantum computation and
processing, which has been the subject of intense research.
\cite{Flatte:book,Imamoglu:4204}

At the same time, recent advances in fabrication, manipulation and
probing techniques of semiconductor quantum dots (QDs) have allowed
several groups to successfully manipulate coherently the exciton
population of a single QD
\cite{Stievater:133603,Kamada:246401,Htoon:087401,Zrenner:612,Wang:035306,Stufler:121301}
using strong resonant laser pulses and different probing techniques.
In particular, Zrenner \textit{et al.} have developed a single
self-assembled QD photodiode \cite{Zrenner:612} in which {\em
coherent} population inversion is induced by a strong and carefully
tuned optical pulse and probed by the photocurrent signal. In their
device the resonant pulse generates an electron-hole pair in the QD,
and with the help of an external gate voltage, the carriers can
tunnel into nearby contacts. This process generates the photocurrent
signal used to monitor the coherent state of the system.

In this paper we model the dynamics of a self-assembled QD
photodiode in the presence of {\em elliptically polarized} laser
pulses. We use a density matrix approach that incorporates dipole
coupling, multi-exciton states, and dephasing mechanisms introduced
by parasitic levels in the structure (associated with the wetting
layer). The electron and hole tunneling processes are introduced via
rates from a microscopic model of the structure, which includes the
electron-hole interaction. \cite{Villas-Boas:057404} We show that
for given elliptical polarization of light we can produce a {\em
spin-polarization reversal} in the photocurrent signal as we
increase the pulse area -- an effect that can be used as a dynamical
spin switch of the generated photocurrent. Moreover, we show that
varying intensity of even \emph{circularly} polarized light can
control the spin polarization of the photocurrent for large
anisotropic exchange; the effect persists in realistic QD ensembles,
allowing the generation of intense spin-polarized current.

\begin{figure}[tbh]
\includegraphics*[width=1.0\linewidth]{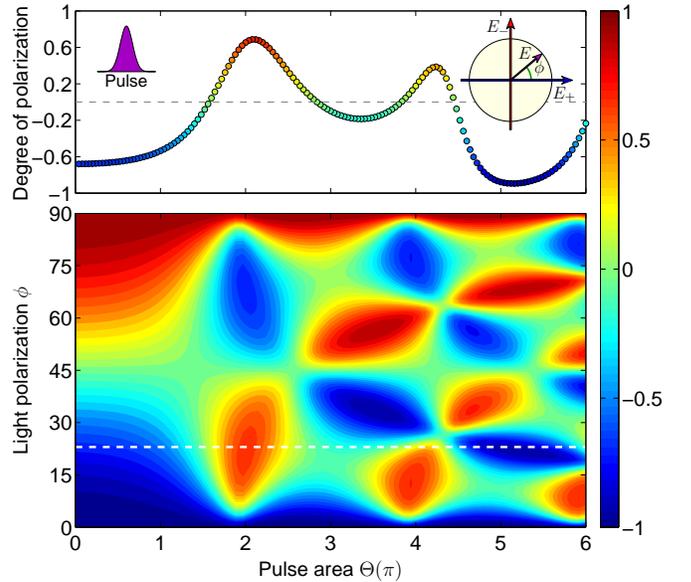}
\caption{(Color online) Bottom panel: degree of spin polarization of
the photocurrent signal as function of pulse area $\Theta$ and
polarization angle of the incoming light. Upper panel: results for
$\phi=\pi/8\simeq23^o$ (dashed white line on the contour plot) where
local maximum contrast is evident. Notice spin reversal of the
photocurrent signal for $\Theta \simeq 2\pi$. Right and left insets
show light polarization angle and the 0.5 ps FWHM Gaussian pulse
used in the simulation, respectively.} \label{fig1}
\end{figure}

An example of those effects can be seen in Fig.\ \ref{fig1}, showing
a contour plot of the degree of spin polarization of the
photocurrent signal in a single QD as function of the pulse area and
polarization angle $\phi$ of the laser pulse. The upper panel shows
the result for $\phi=\pi/8\simeq23^o$, where we have high contrast
in the spin-polarization reversal. We emphasize that this spin
modulation of the photocurrent is achieved for {\em constant}
elliptical polarization, i.e., constant ratio $E_-/E_+$, as defined
by the angle $\phi$, while changing only pulse area (intensity).

\begin{figure}[tbh]
\includegraphics*[width=1.0\linewidth]{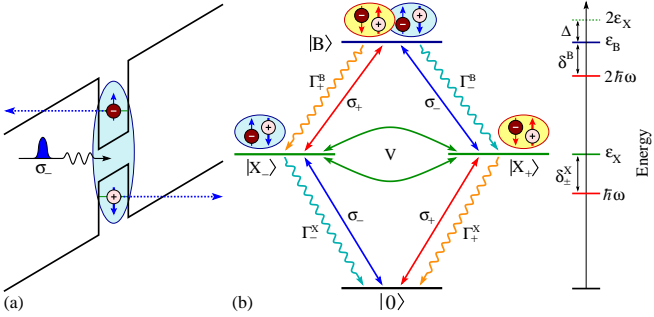}
\caption{(Color online) Schematic band structure and level
configuration of a single QD photodiode in the presence of pulsed
polarized light. (a) A polarized laser pulse creates excitations
(exciton, biexciton) in the dot, and an applied gate voltage forces
the electrons and holes to tunnel out, generating the measured
photocurrent signal. (b) Processes and levels involved in this
system: $|\mathrm{X}_\pm\rangle$ and $|\mathrm{B}\rangle$ are the
different exciton polarizations and biexciton states.} \label{fig2}
\end{figure}

To predict this novel behavior, we have used a realistic
model of the system based on the excitonic levels of a single
quantum dot, as summarized in Fig.\ \ref{fig2}.  Using a unitary
transformation \cite{Villas-Boas:125342} to remove the fast
time-dependent part, one can write
\begin{eqnarray}
H &=&\delta^\mathrm{X}_{\pm}|\mathrm{X}_\pm\rangle \langle
\mathrm{X}_\pm|+\delta^{\mathrm{B}}|\mathrm{B}\rangle \langle
\mathrm{B}|-\frac{1}{2}\Bigl[\Omega^\mathrm{X}_\pm(t)|0\rangle
\langle \mathrm{X}_\pm| \nonumber \\
&&+\Omega^\mathrm{B}_\pm(t)|\mathrm{X}_\mp\rangle \langle
\mathrm{B}| -2V|\mathrm{X}_-\rangle \langle \mathrm{X}_+|
+h.c.\Bigr]. \label{eq1}
\end{eqnarray}
Here, $X_\pm$ are the excitons with total spins $\pm1/2$,
$\delta^\mathrm{X}_{\pm}=\varepsilon_\mathrm{X_{\pm}}- \hbar\omega$
accounts for the detuning of the exciton from the laser energy
$\hbar \omega$,
$\delta_\mathrm{B}=\varepsilon_\mathrm{B}-2\hbar\omega$ is the two
photon biexciton detuning, and $\Omega^\mathrm{X}_\pm(t)=\langle 0|
\vec{\mu}\cdot \vec{E}_\pm(t)|\mathrm{X}_\pm\rangle/\hbar$,
$\Omega^\mathrm{B}_\pm(t)=\langle \mathrm{X}_\mp| \vec{\mu}\cdot
\vec{E}_\pm(t)|\mathrm{B}\rangle/\hbar$, are the optical matrix
elements, where  $\vec{\mu}$ is the electric dipole moment which
couples the excitonic transition to the polarization component
$\vec{E}_\pm(t)$ of the radiation field.  The quantity $V$ denotes
the anisotropic electron-hole (\textit{e-h}) exchange interaction
which originates either from shape (in III-V materials) or crystal
(in II-VI materials) anisotropies of the QD.  $V$ provides mixing
between spin-defined excitons $X_\pm$, which are directly produced
by $\sigma_\pm$ polarized light.
\cite{Gammon:3005,Bonadeo:1473,Bayer:1748}

To obtain the dynamics of the full system we use a density matrix
formalism of the form
\begin{equation}
\frac{d\rho}{dt} = -\frac{i}{\hbar}[H, \rho] + L(\rho),
\label{eqrho}
\end{equation}
where the first term on the right yields the unitary evolution of
the quantum system and $L(\rho)$ is the dissipative part of the
evolution; we assume the Markovian approximation. Our model
considers two types of population decay, one due to the spontaneous
decay given by the recombination rate
$\Gamma_\pm^\mathrm{\{X,B\}_{rec}}$, and another due to electron or
hole tunneling into the contacts, with rates
$\Gamma_\pm^\mathrm{\{X,B\}_{tun}}$. The total population decay rate
for each individual channel can then be written as
$\Gamma_\pm^\mathrm{X,B}=\Gamma_\pm^\mathrm{\{X,B\}_{tun}}
+\Gamma_\pm^\mathrm{\{X,B\}_{rec}}$.

Our simulation uses parameters obtained either from experiments or
from realistic estimates. We assume exciton resonant excitation, so
that $\delta^\mathrm{X}_{\pm}=0$, while $\delta^\mathrm{B}=\Delta=3$
meV is the biexciton binding energy. \cite{Zrenner:612} The
tunneling rates $\Gamma_\pm^\mathrm{\{X,B\}_{tun}}$ are estimated
using a microscopic model, \cite{Villas-Boas:057404} as function of
the external gate voltage, and are in agreement with experimental
values; for example, the exciton tunneling time is found to be
$\Gamma_\pm^X \simeq 12$ ps. \cite{Zrenner:612}

For InGaAs self-assembled QD samples, the \textit{e-h} exchange
interaction is typically a few tens of $\mu$eV, equivalent to an
oscillation time between $X_\pm$ states of the order of tens of ps.
This time is much shorter than the recombination time ($\simeq 1$
ns) so that the QD ``visits" both states many times before the
exciton recombines. That is the reason why it is so difficult to see
polarization of the photoluminescence coming from neutral excitons
without the use of a magnetic field, which restores the spin degree
of freedom as a good quantum number. \cite{Kroutvar:81} In contrast,
a photocurrent measurement depends on the tunneling time, which is
controllable by an external gate voltage and can be tuned from a few
picoseconds and higher. As such, it should be possible to observe
spin polarization in the outgoing photocurrent for short tunneling
times. For example, for a purely circularly polarized pulse we
should be able to produce almost 100\% polarized photocurrent just
because the readout and manipulation times in this case would be
much faster (few picoseconds) than the time for the system to visit
the other polarized state (tens of picoseconds) or any other spin
dephasing mechanism (usually in the nanosecond scale). It is
important to mention that the polarization of nuclear spins in this
kind of QD can be an important source of spin dephasing;
\cite{Tartakovskii:057401,Braun:116601,Bracker:047402} however, the
time scale involved in these processes is also much longer than the
photocurrent measurement.

Our model also includes the leakage to wetting layers (WL) states,
as described in previous work. \cite{Villas-Boas:057404} This is an
important mechanism for the dephasing of the charge state (exciton),
but it does {\em not} affect directly the ratio between the
different polarizations of the photocurrent, since the WL states are
not spin dependent; as such, their effect turns out to be not as
important in the behavior reported here.

Knowing the tunneling rates, the photocurrent signal can be easily
computed by integrating in time over all possible channels/modes of
particles that tunnel out. As the system is back in the vacuum state
$|0\rangle$ when the next pulse arrives, we can write each
individual electron spin-components of the photocurrent as
\begin{equation}
I_\mp=fq\int_{-\infty}^\infty\left[\Gamma_{\pm}^{B_\mathrm{tun}}
\rho_{\mathrm{BB}}(t)
+\Gamma_{\pm}^{\mathrm{X_{tun}}}\rho_{\mathrm{X_\pm
X_\pm}}(t)\right]dt, \label{eq2}
\end{equation}
where $f$ is the pulse repetition frequency (we use $f=82$ MHz as in
Zrenner's experiment \cite{Zrenner:612}) and $q$ is the electronic
charge. The occupation of each state can be obtained by numerically
solving Eq.\ (\ref{eqrho}), and then integrating to give each
component of the polarized photocurrent (\ref{eq2}). The resulting
spin-polarized photocurrent signal can then be computed by $P=(I_+
-I_-)/(I_+ + I_-)$. Notice that the spin polarization of the current
is associated with the spin of the electron only, as the hole is
known to have a rather high spin flip rate due to spin-orbit
interaction.  As such, the hole loses its spin memory soon after
tunneling and only the electron spin remains.

Figure \ref{fig1} shows a contour plot of typical results as
function of the total pulse area
$\Theta=\int_{-\infty}^{\infty}\overrightarrow{\mu}\cdot
\overrightarrow{E}(t)dt/\hbar$. Here,
$\overrightarrow{E}(t)=\overrightarrow{E}_+(t)+\overrightarrow{E}_-(t)$
is the total pulse amplitude with polarization components
$\vec{E}_\pm(t)$ and polarization angle $\phi= \tan ^{-1} (E_-/E_+)$
of the incoming pulse of Gaussian shape and 0.5 ps FWHM. We can see
a rich dependence with polarization angle $\phi$, where 0
corresponds to a $\sigma_+$, and 90$^o$ to a $\sigma_-$ pulse as can
be seen in the top right inset of Fig.\ \ref{fig1}. In detail, the
upper panel shows results for $\phi=\pi/8\simeq23^o$, where we can
see a maximum contrast in the spin-polarization reversal, going from
$\simeq 65 \%$ spin down for low laser intensity to $\simeq 65\%$
spin up for pulse area $\Theta \simeq 2\pi$.  Notice this large
change in photocurrent polarization arises due to the differences in
state populations, as controlled by the laser intensity (pulse
area), despite the fact that the system is pumped with a
\emph{constant} ratio $E_-/E_+$, as defined by the angle $\phi$.

\begin{figure}[tbh]
\includegraphics*[width=1.0\linewidth]{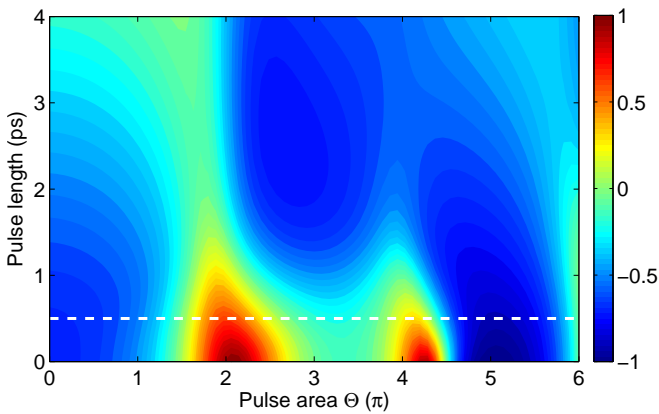}
\caption{(Color online) Photocurrent polarization map as function of
pulse area $\Theta$ and pulse length for elliptically polarized
incoming light with phase $\phi=\pi/8\simeq23^o$, corresponding to
the dashed line in Fig.\ \ref{fig1}. Dashed line here shows pulse
length used in Fig.\ \ref{fig1}. The strong inversion of the signal
is suppressed for longer pulses, and the remaining features are due
to the small anisotropic \textit{e-h} exchange interaction ($V=70
\,\mu$eV).} \label{fig3}
\end{figure}

To obtain a more complete picture of this behavior, Fig.\ \ref{fig3}
shows a contour plot of the polarization signal for the optimal
angle $\phi\simeq23^o$ as function of the pulse duration and pulse
area. Notice that the spin reversal with $\Theta$ occurs only for
short pulses. We know from our previous studies that longer pulses
reduce the occupation probability of the biexciton state,
\cite{Villas-Boas:057404} which in turn reflected in the
photocurrent polarization signal exhibiting smaller or no switching
with larger pulse area. For pulses longer than 2 ps the biexciton
occupation is negligible, and the features seen in Fig.\ \ref{fig3}
are caused only by the anisotropic \textit{e-h} exchange interaction
which we assume to be 70 $\mu$eV, as in the experiment by Muller
\textit{et al.} \cite{Muller:981} For longer pulses, the time to
visit the other spin state starts to become important; without the
anisotropic \textit{e-h} exchange interaction no polarization change
is seen in Fig.\ \ref{fig3} for pulse widths longer than 2 ps (data
not shown). This demonstrates that the biexciton plays an essential
role in this spin switch device. The biexciton works as an effective
exchange channel between the two polarizations due to nonlinear
optical interference. In fact, neglecting the biexciton state in the
simulation, Fig.\ \ref{fig1} changes its color uniformly from bottom
to top (increasing angle $\phi$) and is basically constant from left
to right (increasing laser intensity; data not shown). In that case,
the degree of polarization would naturally be proportional to the
ratio $E_-/E_+$. \cite{Vasko:073305} In the same way, the analogue
of Fig.\ \ref{fig3} would be of a single color, with the tone being
set by the ratio $E_-/E_+$.

\begin{figure}[tbh]
\includegraphics*[width=1.0\linewidth]{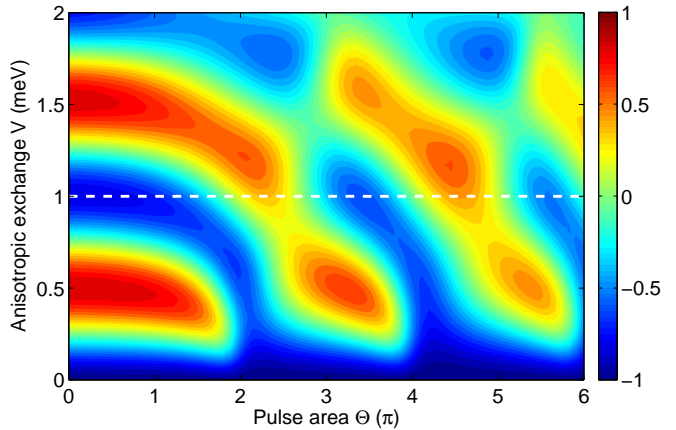}
\caption{(Color online) Photocurrent polarization map vs.\ pulse
area and anisotropic \textit{e-h} exchange interaction $V$ for 1 ps
{\em circularly polarized} pulse $\sigma_+$ ($\phi =0$). For strong
exchange interaction $V=1$ meV, dashed line, one can see strong
oscillations of the polarization as the pulse area $\Theta$
increases.} \label{fig4}
\end{figure}

It is interesting to note that a stronger anisotropic \textit{e-h}
exchange interaction can also result in strong spin switching, even
with no elliptical polarization of light. The possibility for such
behavior resides in group II-VI materials with strong crystal
anisotropy or in large shape anisotropies in typical QDs (such as
elongated dots in one direction, even in III-V materials). In that
case, a spin switch device develops as well, as one can see in Fig.\
\ref{fig4}. There we show the degree of photocurrent polarization in
a QD as function of exchange interaction $V$ and pulse area $\Theta$
achieved by applying a {\em circularly polarized} $\sigma_+$ pulse
with FWHM of 1 ps. We notice, for example, that for a QD with $V=1$
meV, as those in Ref.\ [\onlinecite{Finley:153316}], one can achieve
the same type of oscillation in the polarization signal of the
photocurrent as in the case of the shorter elliptically polarized
pulse in a typical (small $V$) QD.

Most interestingly, it is also possible to observe the predicted
effect in QD \emph{ensembles}. The spin-polarized photocurrent
response in ensembles would not only make its experimental
realization much easier, but it will also open the door for
innumerable applications. In an ensemble of QDs we have to address
the consequences of their size distribution: QDs with different
sizes have different energy levels, so that a laser pulse in
resonance with the exciton energy in one dot is detuned with respect
to the neighbors. Different sizes also result in different dipole
moments, so that the neighbor dots will also oscillate with slightly
different Rabi frequency. Another important issue is the shape
anisotropy, which would result in different anisotropic \textit{e-h}
exchange interaction.  It is known that $V$ can change dramatically
(even by $100\%$) over the dot population, possibly the result of
different dot environments, as well as shapes and sizes.

To model the ensemble we then assume a Gaussian distribution of QD
sizes such that the photoluminescence of the exciton state would
have a broadening of 20 meV, which indeed is a highly homogeneous
ensemble, but still feasible in experiments.
\cite{Malik:1987,Raymond:187402} Directly dependent on that we also
assume that the dots have a 20\% change in their dipole moment, like
in the model used by Borri \textit{et al.} \cite{Borri:081306} to
describe the Rabi oscillations in an ensemble of dots. The degree of
polarization for the ensemble can then be obtained by integrating
the contribution of individual dots as $P_\mathrm{ens}=\int_0^\infty
\int_0^\infty P(\delta(R),\Omega(R),V)f(R)g(V)dR\,dV$, where $f$ and
$g$ describe the corresponding size and $V$ distribution functions.
We assume that the detuning and the difference in dipole moments are
correlated (as per dot size), while they are uncorrelated with
respect to the anisotropic \textit{e-h} exchange interaction (for
which we also assume a Gaussian distribution $g(V)$ with FWHM of
70$\mu$eV around the typical value $V=70\mu$eV).

\begin{figure}[tb]
\includegraphics*[width=1.0\linewidth]{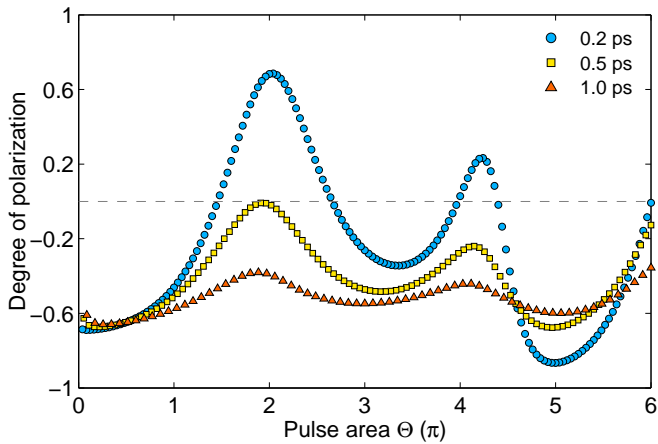}
\caption{(Color online) Polarization of the photocurrent signal as
function of the pulse area for an ensemble of QDs for three
different pulse widths and polarization angle $\phi\simeq23^o$.  The
Gaussian size distribution has 20\% FWHM.} \label{fig5}
\end{figure}

In Fig.\ \ref{fig5} we show the results for three different pulse
widths. We can see that the size distribution tends to suppress the
effect, as one would expect, but there is a non-zero polarization
even for small pulse area. In particular, compare the middle trace
here (squares) for 0.5 ps pulse, with the upper panel in Fig.\
\ref{fig1}. Notice that longer pulses (triangles) exhibit an
oscillatory polarization with pulse area, but no signal inversion.
However, shorter pulses (circles) restore a great deal of modulation
and polarization reversal with $\Theta$, {\em despite QD ensemble
inhomogeneities}. The polarized signal has a strong dependence on
the pulse area (for fixed pulse duration), and at around $2 \pi$
there is a {\em reversal} in the signal polarization. This is an
important and exciting result which reflects nonlinear effects
clearly involving the biexciton state. Being able to dynamically
invert the spin photocurrent with pumping intensity can prove very
useful in controlling the polarization in QDs and associated
devices.

We have presented a model to describe non-linear spin-dependent
effects in self-assembled QD photodiodes. Our model includes
excitons with different polarization and biexciton states, and is
based on the density matrix formalism. We show that for an
elliptically polarized pulse the photocurrent signal presents a
clear inversion of its electron spin component as function of the
pulse area for a single QD. We also show that such effect can still
be observed in an typical ensemble of dots, which makes it more
attractive for device applications.

This work was supported by the Indiana 21st Century Fund, and by
the BNNT project at OU.

\end{document}